\documentclass[letterpaper,twocolumn,aps,prl,showpacs,floatfix]{revtex4}

\usepackage{graphicx}
\usepackage[latin1]{inputenc}
\usepackage{color}
\usepackage{amsmath,amssymb,amsfonts}
\usepackage{subfigure}

\begin{document}
\title{Evolution of the potential-energy surface of amorphous silicon}

\author{\firstname{Houssem} \surname{Kallel}}
   \email{houssem.kallel@umontreal.ca}
   \affiliation{D\'{e}partement de Physique and Regroupement Qu\'{e}b\'{e}cois
   sur les Mat\'{e}riaux de Pointe (RQMP), Universit\'{e} de Montr\'{e}al, C.P.
   6128, Succursale Centre-Ville,\\Montr\'{e}al, Qu\'{e}bec, Canada H3C 3J7}

\author{\firstname{Normand} \surname{Mousseau}}
   \email{normand.mousseau@umontreal.ca}
   \affiliation{D\'{e}partement de Physique and Regroupement Qu\'{e}b\'{e}cois
   sur les Mat\'{e}riaux de Pointe (RQMP), Universit\'{e} de Montr\'{e}al, C.P.
   6128, Succursale Centre-Ville,\\Montr\'{e}al, Qu\'{e}bec, Canada H3C 3J7}

\author{\firstname{Fran\c{c}ois} \surname{Schiettekatte}}
   \email{francois.schiettekatte@umontreal.ca}
   \affiliation{D\'{e}partement de Physique and Regroupement Qu\'{e}b\'{e}cois
  sur les Mat\'{e}riaux de Pointe (RQMP), Universit\'{e} de Montr\'{e}al, C.P.
6128, Succursale Centre-Ville,\\Montr\'{e}al, Qu\'{e}bec, Canada H3C 3J7}

\begin{abstract}

The link between the energy surface of bulk systems and their dynamical
properties is generally difficult to establish. Using the
activation-relaxation technique (ART nouveau), we follow the change in the
barrier distribution of a model of amorphous silicon as a function of the
degree of relaxation. We find that while the barrier-height distribution,
calculated from the initial minimum, is a unique function that depends only
on the level of distribution, the reverse-barrier height distribution,
calculated from the final state, is independent of the relaxation, following a
different  function. Moreover, the resulting gained or released
energy distribution is a simple convolution of these two distributions
indicating that the activation and relaxation parts of a the elementary
relaxation mechanism are completely independent. This characterized energy
landscape can be used to explain nano-calorimetry measurements.

\end{abstract}

\pacs{10.4}
\maketitle

The concept of energy landscape has been used extensively in the last decade
to characterize the properties of complex materials. In finite systems, such
as clusters and proteins, the classification of energy minima and energy
barriers has shown that it is possible to understand, using single-dimension
representations such as the disconnectivity graph, the fundamental origin of
cluster dynamics and protein folding~\cite{wales_book}. For bulk systems,
where the relations are more complex, the energy landscape picture has helped
further our qualitative understanding of glassy dynamics and of the evolution
of supercooled liquids through their inherent structures~\cite{wales_book}.
Similarly, applied to amorphous semiconductors, it has revealed an unexpected
simplicity that suggests universality~\cite{barkema98,middleton01,
valiquette03}. In spite of these contributions, many questions remain
regarding the evolution of the local structure of the energy landscape itself
as a function of relaxation, an evolution that determines the dynamics of
disordered materials away from equilibrium~\cite{karmouch07}. For example,
following measurements of heat released during the annealing of damage
generated by low-energy ion implantation in amorphous silicon, it was
suggested that the activation energy and the amount of heat released are
uncorrelated~\cite{mercure05,karmouch07}, microscopic details were still
missing, however, as well as a general picture of the phenomenon. Results
presented here tie experiments and simulations and provide a clear link
between the structure of the energy landscape and experimental observations,
supporting the importance of this concept.

More precisely, we focus on amorphous silicon, a model system studied
extensively, both experimentally and theoretically, over the
years~\cite{drabold09, karmouch07,valiquette03}. Using ART
nouveau~\cite{malek00}, a saddle-point finding method, we characterize the
evolution of the local energy landscape, defined by the distribution of
transition states and adjacent energy minima around a local configuration, as
a function of relaxation. The resulting theoretical picture can be used to
understand and explain recent nano-calorimetric measurements on ion-implanted
\emph{a}-Si samples~\cite{mercure05,karmouch07}. 

\begin{figure}[t]
	\centering
		\includegraphics[height=6cm]{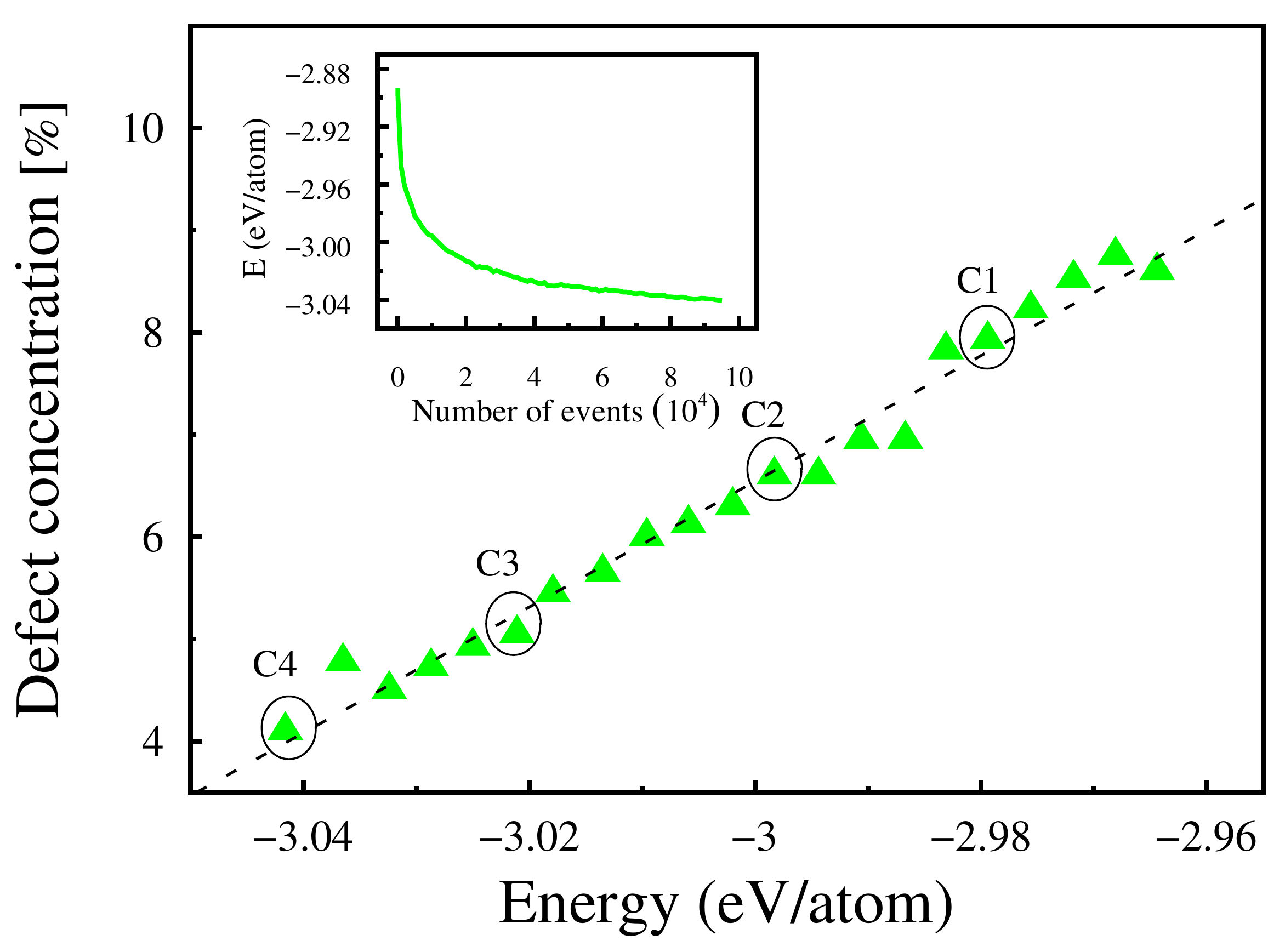}
	\caption{ Fraction of coordination defects (mostly 3-fold and 5-fold coordinated atoms) as a function of the configurational energy per atom (in eV) for the 4000-atom model of \emph{a}-Si. The large circles indicate the location of the four configurations selected for further analysis. Inset: energy per atom as a function of event during the preparation of the configurations.}
	\label{fig:relaxation}
\end{figure}

Our \emph{a}-Si model is relaxed using the Stillinger-Weber potential with
parameters adjusted specifically to recover the structural and vibrational
properties of the amorphous system~\cite{vink01}. While this potential has a
number of shortcomings, its structural evolution as a function of relaxation
is in good qualitative agreement with experiment as discussed below. Moreover,
it is fast enough to allow for an extensive search of transition states,
necessary to construct reliable energy distributions. Starting from a
4000-atom random configuration in a cubic box with periodic-boundary
conditions, the system is relaxed using ART nouveau: beginning in a local
energy minimum, the system is brought to an adjacent transition state and
relaxed into a new minimum. The move is accepted using a Metropolis criterion
with a temperature of 0.25 eV, following Ref.~\cite{malek00,valiquette03}. A
total of 95~445 events is necessary to relax the model from an initial energy
per atom of -2.89 eV/atom going down to -3.04 eV/atom (see
Fig.~\ref{fig:relaxation} in inset). Figure ~\ref{fig:relaxation} shows the
correlation between the energy per atom and the bond-defect concentration
(mostly three-fold and five-fold coordinated Si). Following this relaxation,
we find a linear relation between the width of the TO peak, calculated using
the relation of Ref.~\cite{vink02}, and the configurational energy, with a
slope of 0.989 (KJ/mol)/deg, 
in good agreement with experiment~\cite{roorda91,mercure05}.

\begin{table}[tb]
\caption{Properties of the four sampled 4000-atom \emph{a}-Si configurations including energy per atom and the fitting parameters to Equation ~\ref{eq1} for the forward-barrier (FB) distribution.}
\begin{center}
\begin{tabular}{c|cccc}
		& C1 & C2 & C3 & C4 \\ \hline
		Energy/atom   & -2.980& -2.996&-3.019&-3.042  \\
	$\langle E_{FB}\rangle$ (eV) &1.919& 2.190 & 2.430 & 2.587 \\
	$\sigma_{FB}$ (eV)        & 1.178 & 1.153 & 1.141 & 1.164\\
	    A (eV$^{-2}$) &0.17&0.16&0.14&0.13\\
	\end{tabular}
	\end{center}
	\label{tab:forward}
\end{table}

To characterize the evolution of the local energy landscape as a function of
the degree of relaxation, we explore extensively the transition state distributions around four configurations with decreasing internal stress.
The energy per atom for these configurations is indicated in
Table~\ref{tab:forward}. From each of these four generic configurations, we
generate 100~000 events (exit pathways), providing a detailed picture of the
local energy landscape associated with various degrees of relaxation. In each
case, we remove the duplicate events leaving about 12~000 different activated
pathways from the initial local minimum to a nearby one for each of these four
configurations, i.e. about 12~000 different transition points and adjacent
minima. The nature of dominant diffusion and relaxation mechanisms in
\emph{a}--Si have already be discussed
elsewhere~\cite{barkema98,song00,valiquette03} and we focus here on the
distribution of forward and reverse barrier heights (measured, respectively,
from the transition state to the initial and final state) and energy asymmetry
(given by the energy difference between the final and initial state), for each
level of relaxation. From previous analysis, we know that although the total
number of events in \emph{a}-Si is between 30 and 60 per atom, the overall
distribution converges relatively quickly~\cite{valiquette03} and we observe
only negligible differences in the shape of the various energy distributions
by including the events generated after 50~000 or 100~000 ART moves.

\begin{figure}[ht]
	\centering
		\includegraphics[height=12cm]{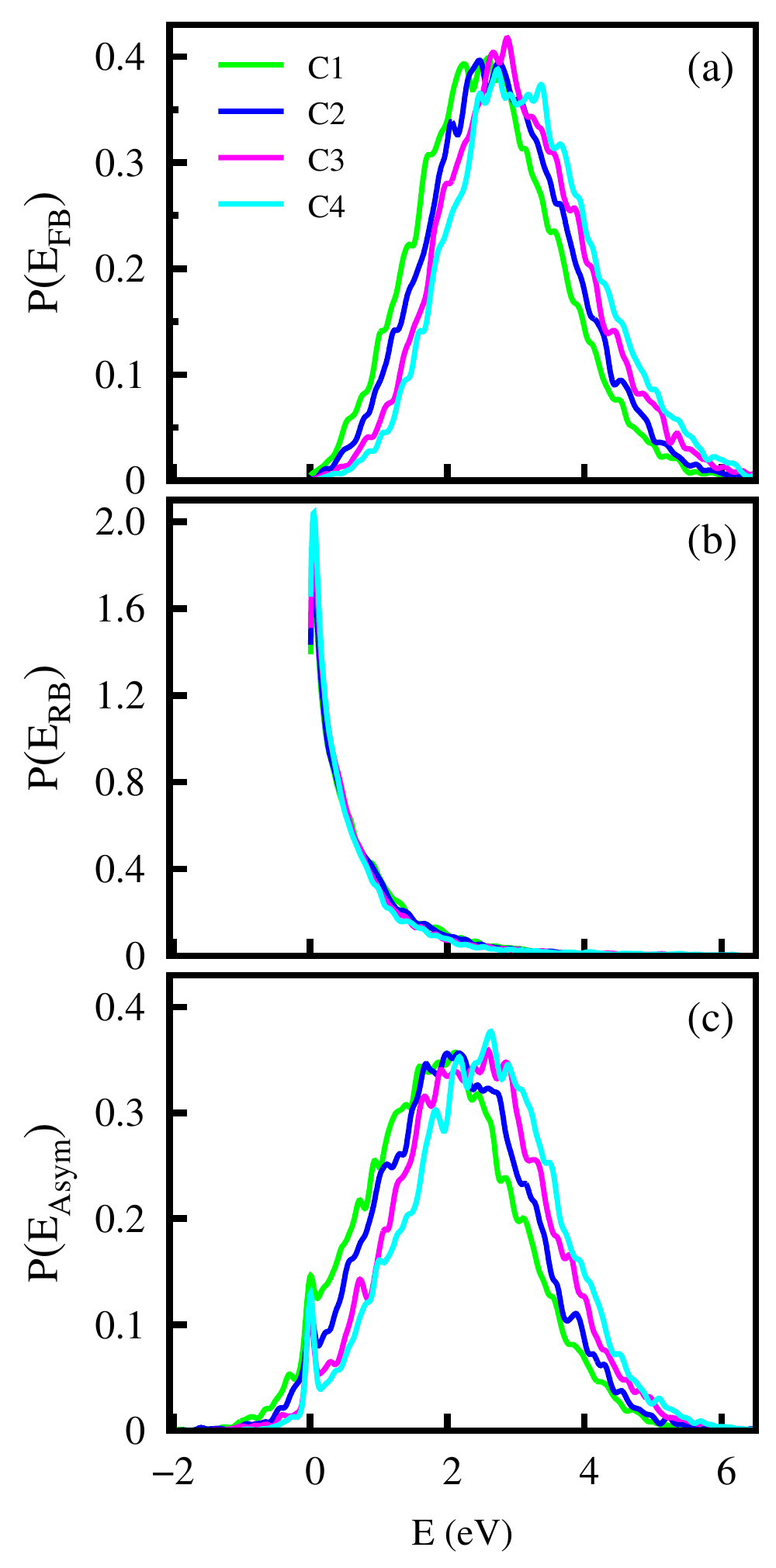}
	\caption{(a) Forward and (b) reverse energy-barrier distribution, and (c) asymmetry energy distribution for configurations C1 to C4.}
	\label{fig:distributions}
\end{figure}

\begin{figure}[ht]
	\centering
		\includegraphics[height=12cm]{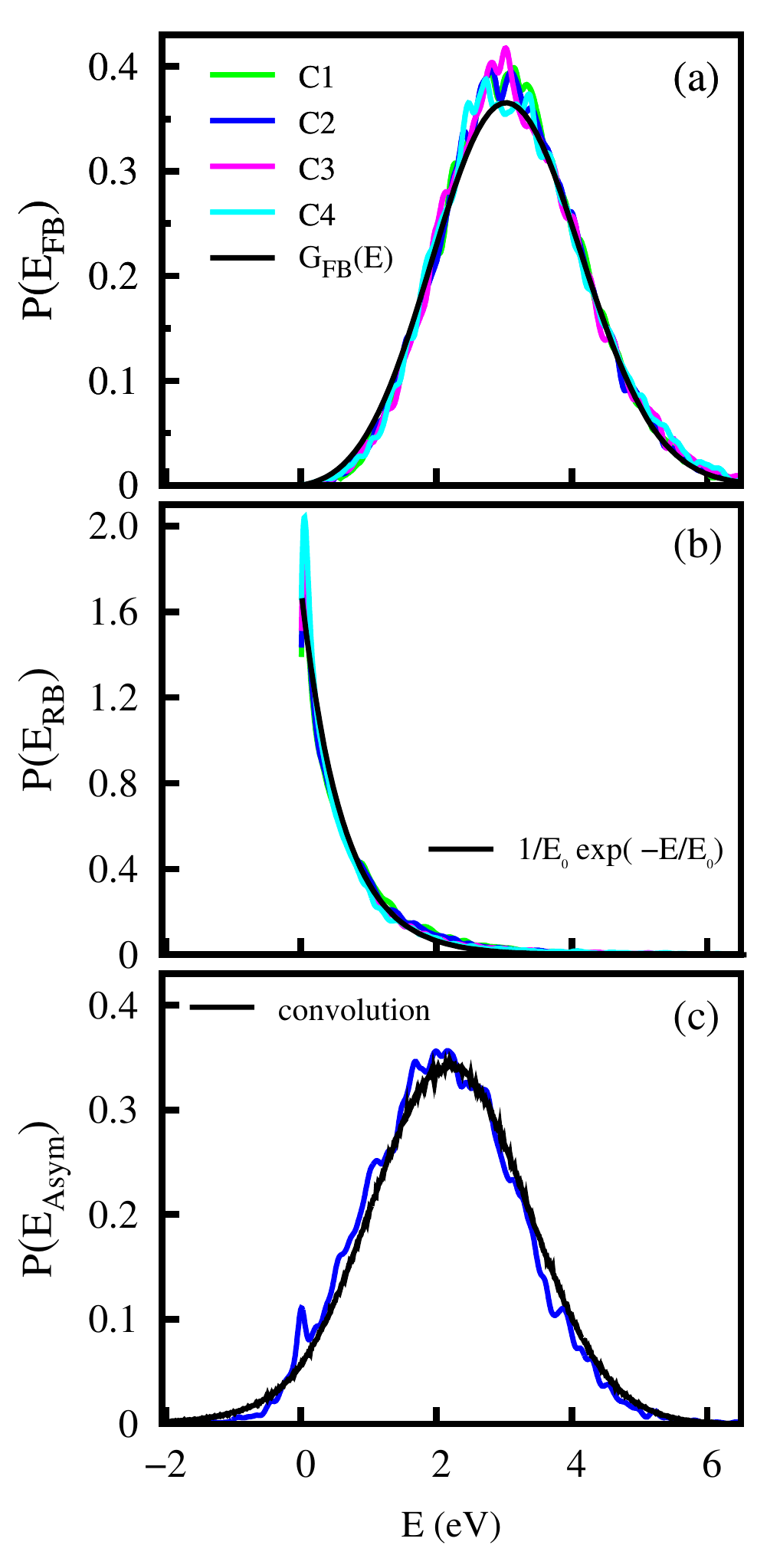}
	\caption{ Collapsed (a) Forward and (b) reverse energy barrier distributions, following the equations given in the text. (c) Asymmetry distribution for configuration C2 and the corresponding distribution generated by convoluting the respective forward and reverse barrier distributions.}
	\label{fig:collapse}
\end{figure}

From transition state theory, we know that the kinetics of
activation-dominated systems is controlled by the the rate of escape from the
initial energy minimum. Figure ~\ref{fig:distributions}(a) shows the distribution of activation barriers from the initial minimum to the transition state (forward barrier or FB) for configurations C1 to C4. We observe that the distribution is continuous and bounded in agreement with previous simulations~\cite{barkema98,valiquette03} but also that it shifts towards higher energy as the configuration are relaxed.  As shown in Fig.~\ref{fig:collapse}(a), however, the four curves are well represented by the same function given by a Gaussian multiplied by the energy:
\begin{equation}
	G_{FB}(E) = A E \exp\left\{ - \frac{(E-\langle E_{FB}^{rel} \rangle)^2}{2 \sigma_{FB}^2}\right\}
	\label{eq1}
\end{equation}
where, as is shown in Table ~\ref{tab:forward}, the peak position controlled
by $E_{FB}^{rel}$ varies with the degree of relaxation but the overall scale
$A$ and the width of the distribution, given by $\sigma_{FB}$, are 
constant. Remarkably, the shift in E$_{FB}^{rel}$ is significantly larger than
the drop in the average strain per atom as the configuration is relaxed:
E$_{FB}^{rel}$ shifts by 0.67 eV from C1 to C4 as the energy per atom
decreases by only 0.06 eV. This result is consistent with the growing
super-Arrhenius relaxation time scale observed in strong glasses and generally
described by the empirically-derived Vogel-Fulcher law~\cite{vf}. However,
while this relation postulates a constant energy-barrier and an unusual
temperature relation, we see here that the reverse is happening: it is the
barrier height that increases with relaxation. The disproportionate shift in
the forward barrier distribution as a function of relaxation can be can be
explained by previous observations that the defected and strained structures
play a major role in the evolution of amorphous systems as diffusion or
rearrangement in perfectly coordinated region of the amorphous network tend to
have an energy cost similar to the of the crystal~\cite{barkema98,drabold09}.

Even though a system's kinetics is conditioned by the barrier height between
minima, the long time evolution of a system is also determined by the overall
event energy asymmetry. For example, if system jumps onto a very energetic
state, with a high energy asymmetry with respect to the initial minimum, it
will come back immediately onto the initial state irrespective of the height
of the forward energy barrier. In effect the presence of these highly
asymmetric states surrounding a local minimum only slows down the relaxation.
To characterize this phenomenon, we plot the reverse energy barrier
distribution (Fig.~\ref{fig:distributions}(b)), i.e., the energy barrier
height measured from the final minimum. As had been observed before, we note
that the forward and reverse energy-barrier distributions are qualitatively
different ~\cite{valiquette03}: instead of a Gaussian, we observe an
exponential distribution with a large proportion of irrelevant states and very
few low-energy minima contributing to the evolution of \emph{a}-Si into a more
stable local state (also observed for Lennard-Jones and \emph{a}-Si in a
slightly different set-up in Ref.~\cite{middleton01,valiquette03}). Contrary
to the behavior of the forward energy distribution, we find that the reverse
energy barrier distribution is essentially independent of the relaxation
level, with the exception of region of very low energy, zero eV where the
identification of recurring events is sometimes tricky. Neglecting this
region, we see that all backward barrier distributions are represented by the
same exponential, $\exp (-\Delta E/E_0)$, where E$_0$ = 0.60 eV (see Fig.
~\ref{fig:collapse}(b)).

These two distributions, forward and backward, cannot be measured directly experimentally. However, the heat released during relaxation is directly associated with the energy asymmetry, i.e., the energy difference between the initial and final minima. Fig. ~\ref{fig:distributions}(c) shows the energy asymmetry distribution surrounding configurations C1 to C4, a structure that changes with the degree of relaxation. While the shape of the distribution looks complex, it is a simple convolution of the two independent forward and backward energy barrier distributions, as is shown in Fig. ~\ref{fig:collapse}(c) . Not only do these two distributions behave differently with the relaxation, these turn out to be independent: the height of a barrier does not provide any information regarding the energy of the final minimum.

The energy landscape picture developed here can be used to explain recent
differential nano-calorimetric measurements made on implanted
\emph{a}--Si~\cite{mercure05,karmouch07}, offering an experimental check on
these results. Samples are implanted at low temperature and then heated, while
heat released is measured, revealing a number of features: First, the heat
flux released as the sample is heated is relatively uniform as a function of
temperature, for temperatures ranging from 118 to 775 K; second, the
stored-strain, measured as the heat rate, saturates as a function of
implantation fluence; third, for low fluences, the heat-rate follows nearly
the same curve, irrespective of the implantation temperature. For fluences
near saturation, the heat release amplitude increases with decreasing
implantation temperature. These results were explained by the presence of a
continuous distribution of states where each annealing event releases an
amount of heat independent of its activation energy. Dynamic annealing was
considered responsible for the increasing signal with decreasing temperature
near saturation: some configurations are only accessible if underlying, low
activation energy configurations survived dynamic annealing.

The saturation of stored-strain can be readily explained within the context of
energy landscape: We have seen that the barrier-height distribution shifts to
lower energy as the strain level increases. As this shifts occurs, the number
of barriers within k$_{B}$T increases reaching a point, for a given
temperature, at which relaxation takes place faster than the strain
increase, leading to the saturation.

\begin{figure}[ht]
	\centering
	
	\includegraphics[height=6cm]{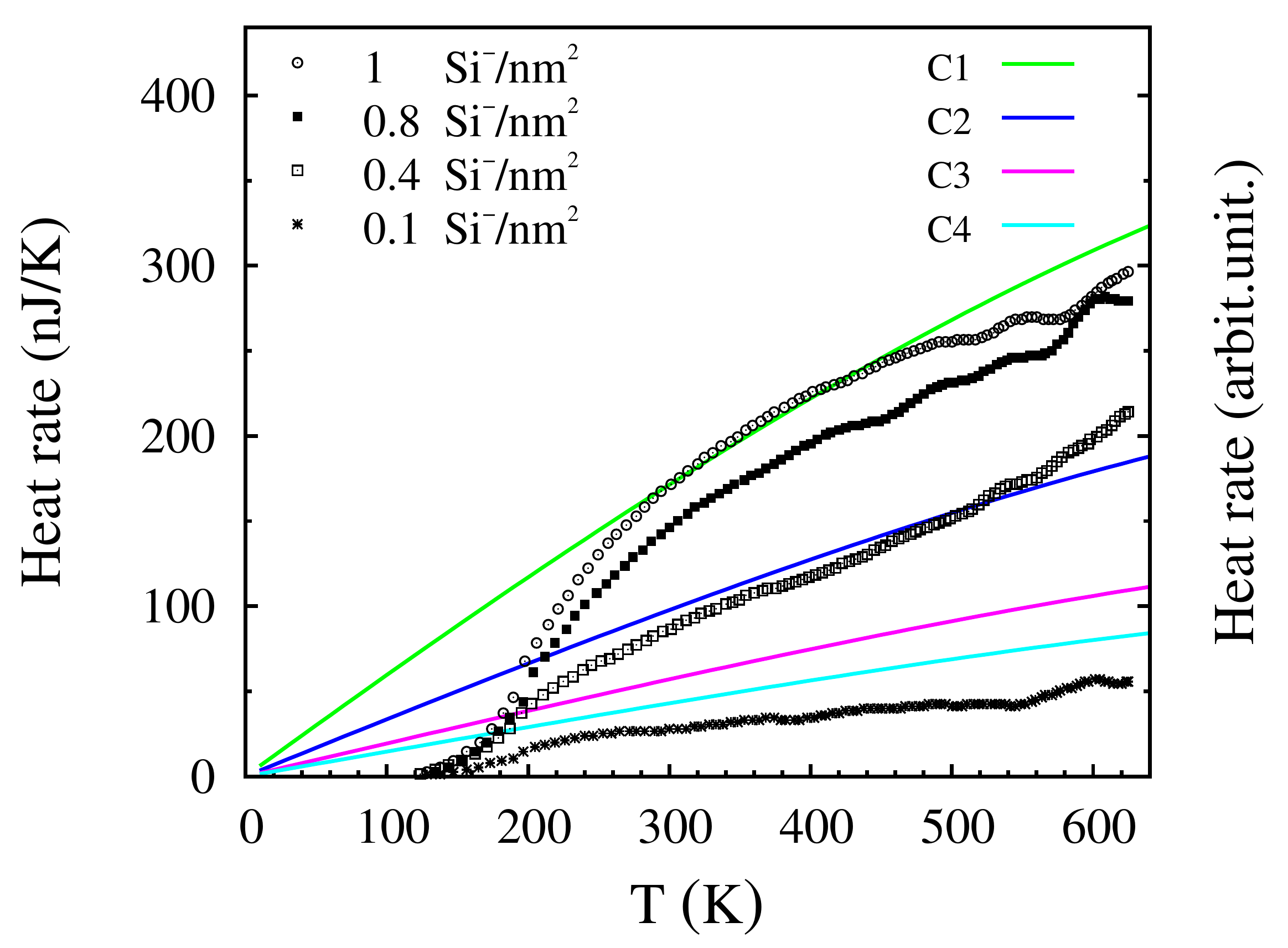}
	\caption{Heat flux as a function of temperature for Configurations C1 to C4 (lines) compared with the heat flux released during relaxation of ion-damaged \emph{a}-Si samples  (experimental data from Karmouch \emph{et al.}~\cite{karmouch07}).}
	\label{fig:flux}
\end{figure}

Understanding the behavior of the released heat-flux requires more analysis.
For this, we suppose that since the energy-barrier distribution shifts rapidly
to higher energy as the system is relaxed, low-energy barriers present in the
initial configuration, before heating is started, are not replaced: once a
barrier is ``used'' it do not reappear at once because heating is relaxing the
system. We do not have to worry about the change in shape at high-energy
because these events have an exponentially small probability of occurring
during the fast heating. Because the reverse distribution is independent of
the relaxation state, we also pose that it remains constant during the heating
process.

Starting from one initial minimum, we assign a temperature-dependent Boltzmann
probability of crossing each barrier, posing a constant attempt
frequency~\cite{song00}. Using a constant heating rate, we select
forward-energy barriers with the Bolztmann-weighted probability times the
probability distribution and the reverse barrier with the exponential
distribution. An event is accepted only if the final state has an energy lower
than the initial state, since high-energy states will be unstable and will
jump back immediately to the original minimum. The resulting heat flux, shown
in Fig.~\ref{fig:flux}, compares very well with the experimental results
allowing us to relate this behavior to two factors: first, the number of
barriers being crossed as a function of temperature increases linearly only
(because $T$ is very small with respect to $E_{FB}$) and most barriers crossed
are of the order of a few k$_{B}$T. However, since most of the weight in the
reverse-barrier distribution is concentrated near zero, the number of crossing
leading to heat released does not change significantly, leading to the
experimentally observed uniform increase as a function of temperature. We can
see in Fig.~\ref{fig:flux}, however, that the exponential fall in reverse
barrier flattens the heat rate curve for well-relaxed sampled, corresponding
to very low fluence, in agreement with experiments.

In conclusion, a detailed analysis of the evolution of the energy landscape for model
\emph{a}--Si shows features that provide a clear picture of defect relaxation in support of
recent differential nano-calorimetric measurements. Because the forward and reverse energy
barrier distributions are independent, heat released after ion-implantation is proportional to
the number of barrier crossed, which increases linearly with the temperature, for highly
disordered systems, and flattens out for more relaxed configurations, where the impact of the
exponential reverse barrier distribution becomes more important. Moreover, the rapid shift in
the forward-energy barrier distribution as a function of relaxation also explains the
saturation in stored strain at high fluence. While it remains to be confirmed numerically, the
picture developed here should apply to covalent glasses in general.

\section*{Acknowledgements}

This work was funded in part by the Canada Research Chairs program, the Fonds
qu\'{e}b\'{e}cois de recherche sur la nature et les technologies (FQRNT) and
NSERC. Calculations were done using ressources from the R\'eseau Qu\'ebecois
de Calcul de Haute Performance (RQCHP). H.K. is grateful to the Tunisian
Government for a scholarship.


\begin{references}

\bibitem{wales_book} David J.Wales, Energy Landscapes. Applications to Clusters, Biomolecules and Glasses( Cambridge Molecular science, Cambridge, 2004).

\bibitem{barkema98} G. T. Barkema and N. Mousseau, Phys.\ Rev.\ Lett. {\bf 81}, 1865 (1998).

\bibitem{middleton01} T. F. Middleton and D. J. Wales, Phys.\ Rev.\ B \textbf{64}, 024205 (2001)

\bibitem{valiquette03} F. Valiquette and N.Mousseau, Phys.\ Rev.\ B {\bf 68}, 125209 (2003).

\bibitem{karmouch07} R. Karmouch \emph{et al.}, Phys.\ Rev.\ B \textbf{75}, 075304 (2007)

\bibitem{mercure05} J.-F. Mercure, R. Karmouch, Y. Anahory, S. Roorda and F. Schiettekatte, Phys.\ Rev.\ B {\bf 71}, 134205 (2005).

\bibitem{drabold09} D. A. Drabold, Eur.\ Phys.\ J.\ B \textbf{68}, 1 (2009).

\bibitem{malek00} R. Malek and N. Mousseau, Phys.\ Rev.\ E {\bf 62}, 7723 (2000).


\bibitem{vink01} R. L. C. Vink et al., J.\ Non-Crystal.\ Sol. {\bf 282}, 248-255 (2001).


\bibitem{vink02} R. L. C. Vink, G.T. Barkema and W.F. van der Weg, Phys.\ Rev.\ B {\bf 63}, 115210 (2001).

\bibitem{roorda91} S. Roorda et al., Phys.\ Rev.\ B {\bf 44}, 3702 (1991).

\bibitem{song00} Y. Song, R. Malek and N. Mousseau, Phys.\ Rev.\ B {\bf 62}, 15680 (2000).


\bibitem{vf} H. Vogel, Phys.\ Z. \textbf{22}, 645 (1921); G. S. Fulcher, J. Am.\ Ceram.\ Soc.\ \textbf{8}, 339 (1925).



\end{references}
\end{document}